\documentclass[aps,showpacs, nofootinbib]{revtex4}

\newcommand{\be}{\begin{equation}}
\newcommand{\ee}{\end{equation}}
\newcommand{\ben}{\begin{eqnarray}}
\newcommand{\een}{\end{eqnarray}}

\newcommand{\cO}{{\cal O}}

\newcommand{\cL}{{\cal L}}

\newcommand{\p}{\partial}
\newcommand{\na}{\nabla}

\newcommand{\trho}{{\tilde \rho}}

\newcommand{\tg}{\tilde g}

\newcommand{\tR}{\tilde R}

\newcommand{\tH}{\tilde H}

\newcommand{\ga}{\gamma}


\begin{document}

\title{Static black holes and strictly static spacetimes in Einstein-Gauss-Bonnet gravity with gauge field}

\author{Marek Rogatko}
\affiliation{Institute of Physics \protect \\
Maria Curie-Sklodowska University \protect \\
20-031 Lublin, pl.~Marii Curie-Sklodowskiej 1, Poland \protect \\
marek.rogat@poczta.umcs.lublin.pl \protect \\
rogat@kft.umcs.lublin.pl}

\date{\today}

\begin{abstract}
We examine strictly static asymptotically flat spacetimes in Einstein-Gauss-Bonnet gravity with $U(1)$-gauge field, revealing that, up to the small curvature corrections,
static conformally flat slices of the spacetime in question are of Minkowski origin. We consider uncharged and charged black hole solutions in the theory showing that
up to the small curvature limit, they are diffeomorphic to Schwarzschild-Tangherlini or Reissner-Nordstr\"om solutions, respectively.
\end{abstract}

\maketitle

\section{Introduction}
Emergence of black holes and gravitational collapse in generalized Einstein theories of gravity
attract much attention last years. These results are closely related to the problem of classification
of non-singular black hole solutions. The pioneering investigations of the mathematical topics
bounded with the black hole equilibrium states were attributed to Israel \cite{isr}. Then they were developed
in Refs. \cite{mil73}-\cite{rob77}. The alternative proof of the black hole uniqueness theorem was established
by Bunting \cite{bun87} and its generalizations \cite{rub88}-\cite{heus}.
Removing the condition of non-degeneracy, the complete classification of vacuum black hole solutions 
as well as Einstein-Maxwell black holes under the condition that all degenerate components of the event
horizon should carry charge of the same signs, was established \cite{chr99}. This assumption was get rid of
in Ref.\cite{chr06}, were the near-horizon geometry of the black hole was investigated.
\par
On the other hand, it turned out that a construction of the uniqueness black hole theorem for stationary axisymmetric
spacetime was far more complicated task \cite{stat}. Nevertheless, the complete proof was devised by
Mazur \cite{maz} and Bunting \cite{bun} (for various aspects of the black hole uniqueness theorem story see 
\cite{book} and references therein).
\par
In the recent years there was also a resurgence of mathematical works concerning black hole
equilibrium states in higher dimensional spacetime. Higher dimensional black objects like black holes,
black rings and black Saturns were widely examined.
The complete classification of $n$-dimensional 
charged black holes both with non-degenerate and 
degenerate component of the event horizon was obtained in Refs.\cite{gib02a}-\cite{ndim}. On the other hand,
some partial results for the nontrivial case of $n$-dimensional rotating black hole uniqueness theorem were 
provided in \cite{nrot}. 
Matter field behaviors  in the spacetime of higher dimensional 
black hole were examined in Ref.\cite{rog12}.
\par
Attempts of unifying all forces of Nature and constructing quantum theory of gravity also triggered
the researches in the realm of the low-energy string theory. They encompass studies of the 
black hole uniqueness theorem in  dilaton gravity, Einstein-Maxwell-axion-dilaton (EMAD)-gravity as well as
supergravities theories \cite{sugra}. On the other hand,
the strictly stationary static vacuum spacetimes in Einstein-Gauss-Bonnet theory were
discussed in \cite{shi13a}.
The Chern-Simons (CS) modified gravity emerging in string theory as an anomaly-canceling 
term in Green-Schwarz mechanism \cite{gre87}, was treated from the point of view of the uniqueness
of black holes. Namely, it was revealed that a static asymptotically flat black hole
solution is unique to be Schwarzschild spacetime in CS-gravity \cite{shi13}, while static black holes
with $U(1)$-gauge are isomorphic to Reissner-Nordstr\"om one \cite{rog13}.  
\par
Black holes are also key ingredients of the AdS/CFT attitude \cite{adscft}, in the context
of a possible matter configuration in AdS spacetime. 
In Ref.\cite{shi12}
it was found that strictly stationary AdS spacetime did not allow for the existence of nontrivial
configurations of complex scalar or form fields, while the generalization of the aforementioned problem,
in EMAD-gravity with negative cosmological constant was accomplished in Ref.\cite{bak13}.
\par
On the other hand, there has been also a renewed interest in theories with higher curvature terms, which
arise naturally in various contexts, e.g., string theory, braneworld physics (see e.g., \cite{br}), holographic
superconductors. Among
the theories which involve higher derivative curvature terms, the so-called Einstein-Gauss-Bonnet
and its Lovelock generalization are of particular interests. Their equations of motion contain
derivatives of the metric of order no higher than the second and therefore it has been proven
to be ghosts free when expanding about flat spacetime, evading any problem with unitarity \cite{bou85}.
\par
Motivated by the aforementioned 
researches we shall consider the problem of strictly static slices (without Killing horizons) in 
Einstein-Gauss-Bonnet gravity
as well as the uniqueness of uncharged and charged black holes in this theory. In our considerations
we confine our attention to small curvature corrections, i.e., 
to the $\cO(\alpha)$-order, where $\alpha$ is a Gauss-Bonnet coupling constant.
\\
The paper is organized as follows. In Sec.II we review the system under considerations. Sec.III will be devoted to the conditions which should be satisfied for
strictly static slices in Einstein-Gauss-Bonnet gravity with $U(1)$-gauge field. In Sec.IV one considers the uniqueness theorem for both uncharged and charged
black holes in the theory in question, restricting our attention to small curvature corrections.
In Sec.V we conclude our investigations.

\section{Einstein-Gauss-Bonnet gravity with $U(1)$ gauge field}
In this section we shall consider Einstein gravity with Gauss-Bonnet term and $U(1)$-gauge field.
It is provided by the action
\be
I = \kappa~\int d^{n+1} x \sqrt{-{g}}~\bigg[ R + {\alpha }~\bigg(
R_{\mu \nu \ga \delta}R^{\mu \nu \ga \delta} - 4~R_{\alpha \beta} R^{\alpha \beta} + R^2 \bigg)
- F_{\mu \nu}~F^{\mu \nu} \bigg],
\label{act}
\ee
where $\alpha$ is the coupling constant of the Gauss-Bonnet term, regarded as the inverse of the string tension. 
It naturally arises as the next leading order of the $\alpha$-expansion of heterotic string theory and is positively defined \cite{bou85}. 
The dimension of this coefficient is $(length)^2$.
In what follows we confine our attention to the case when $\alpha > 0$.
On this account, the field equations obtained by variation of the action (\ref{act}) imply
\ben
R_{\mu \nu} &=& T_{\mu \nu} - g_{\mu \nu}~{T \over n-1} - \alpha~\tH_{\mu \nu} 
-g_{\mu \nu}~{\alpha \over 1-n}~
\bigg(
R_{\sigma \rho \ga \delta}R^{\sigma \rho \ga \delta} - 4~R_{\ga \delta} R^{\ga \delta} + R^2 \bigg),\\
\na_{\mu} F^{\mu \nu} &=& 0,
\een
where we have denoted by $\tH_{\mu \nu}$ the following expression:
\be
\tH_{\mu \nu} = 2~R_{\mu}{}{}^{\alpha \beta \ga}~R_{\nu \alpha \beta \ga}
- 4~R^{\alpha \beta}~R_{\mu \alpha \nu \beta} 
- 4~R_{\mu \alpha}~R^{\alpha}{}{}_\nu + 2~R~R_{\mu \nu}.
\ee 
On the other hand, the energy momentum tensor 
$T_{\mu \nu} = - {\delta S \over \sqrt{- g}~\delta g^{\mu \nu}}$ of matter fields in question yields
\be
T_{\alpha \beta}(F) = 2~F_{\alpha \ga}~F_{\beta}{}^\ga - {1 \over 2}~g_{\alpha \beta}~F_{\mu \nu} F^{\mu \nu}.
\ee
We consider static spacetime in which the asymptotically timelike Killing vector field 
$k_{\alpha} = \big({\p \over \p t }\big)_{\alpha}$ orthogonal to the $n$-dimensional hypersurface $\Sigma$ 
of constant time, is defined. 
The line element of static spacetime subject to the
asymptotically timelike Killing vector field 
$k_{\alpha} $ and $V^{2} = - k_{\mu}k^{\mu}$ is provided by the following relation:
\be
ds^2 = - V^2 dt^2 + g_{i j}dx^{i}dx^{j},
\label{met}
\ee
where $V$ and $g_{i j}$
are independent of the $t$-coordinate as the quantities defined on
the hypersurface $\Sigma$ of constant $t$. Additionally, we suppose that on $\Sigma$
the electromagnetic potential will be of the form $A_{0} = \psi~dt$.
\par
Taking the form of static metric into account, the corresponding
equations of motion yield
\be
V~{}^{(n)}\na_{i}{}^{(n)}\na^{i} V =
T_{00} + {V~T \over n -1} + 4 ~\alpha~{n-3 \over n-1}~{}^{(n)}\na_{i}{}^{(n)}\na_{j} V
~{}^{(n)}G^{ij} - \alpha~{V \over n-1}~\cL_{GB},
\label{eq1}
\ee
\be
{}^{(n)}\na_{i}{}^{(n)}\na^{i} \psi = {1 \over V}{}^{(n)}\na_{i} \psi {}^{(n)}\na^{i} V,
\label{eq2}
\ee
\ben \label{eq3}
{}^{(n+1)} R_{ij} = 
T_{ij} - g_{ij} {T \over n-1} &-& \alpha~
\bigg( 4V^2~{}^{(n)}\na_{i}{}^{(n)}\na_{m}V~{}^{(n)}\na_{j}{}^{(n)}\na^{m}V \\ \nonumber
&-& {4 \over V^2}~{}^{(n)}\na_{i}{}^{(n)}\na_{j}V~{}^{(n)}\na_{m}{}^{(n)}\na^{m}V \bigg) 
- \alpha~\tH_{ij} + g_{ij}~{\alpha \over n-1}~\cL_{GB},
\een
where in the above relations covariant derivative with respect to the
metric tensor $g_{ij}$ is denoted by ${}^{(n)}\na$,
while ${}^{(n)}G_{ij}(g)$ is the Einstein tensor defined on $n$-dimensional hypersurface $\Sigma$. 
\par
Let us assume further
that we take into account the asymptotically flat spacetime,
i.e, the spacetime  
contains a data set
$(\Sigma_{end}, g_{ij}, K_{ij})$ with gauge fields such that 
$\Sigma_{end}$ is diffeomorphic to ${\bf R}^n$ minus a ball and the 
following asymptotic conditions are fulfilled:
\ben
\vert g_{ij}  - \delta_{ij} \vert + r \vert \p_{a}g_{ij} \vert
+ ... + r^k \vert \p_{a_{1}...a_{k}}g_{ij} \vert +
r \vert K_{ij} \vert + ... + r^k \vert \p_{a_{1}...a_{k}}K_{ij} \vert
\le {\cal O}\bigg( {1\over r^{n-4}} \bigg), \\
\vert F_{\alpha \beta} \vert + r \vert \p_{a} F_{\alpha \beta} \vert
+ ... + r^k \vert \p_{a_{1}...a_{k}}F_{\alpha \beta} \vert
\le {\cal O}\bigg( {1 \over r^{n-2}} \bigg).
\een
Consequently, under the above assumptions,
there is a standard coordinates
system in which we have the usual asymptotic decay properties. Namely,
let us assume that each of the quantities
$V,~\psi,$ and $g_{ij}$, has the asymptotic expansion given by
in the form as follows:
\ben \label{lim}
V &\simeq& 1 - {\mu \over r^{n-2}} + \cO \bigg( {1 \over r^{n-3}} \bigg),\\ \label{lim0}
\psi &\simeq& {Q \over r^{n-2}} + \cO \bigg( {1 \over r^{n-4}} \bigg),\\ \label{lim1}
g_{ij} &\simeq& 1 + {2 \over (n-2)~r^{n-2}} + \cO \bigg( {1 \over r^{n-1}} \bigg),
\een
where $Q$ and $\mu$ are constant and represent electric charge and the ADM mass, respectively.

\section{Strictly static spacetime with matter}
To begin with we examine the strictly stationary case, when in static spacetime under considerations
one has no Killing horizons. Having in mind equation (\ref{eq1}) we obtain the following:
\be
{}^{(n)}\na^{i}~\bigg( {}^{(n)}\na_{i} V - 4 ~\alpha~{n-3 \over n-1}~{}^{(n)}\na^{j}V
~{}^{(n)}G_{ij} \bigg) 
= - {\alpha \over n-1}~V~{}^{(n)}\cL_{GB} + {2(n-2) \over V~(n-1)}~{}^{(n)}\na_{a}\psi ~{}^{(n)}\na^{a}\psi,
\label{sss}
\ee
where by ${}^{(n)}\cL_{GB}$ we have denoted
\ben \label{lgb}
{}^{(n)}\cL_{GB} &=& {}^{(n)}R_{ijkl}~{}^{(n)}R^{ijkl} -4~{}^{(n)}R_{ij}~{}^{(n)}R^{ij} + {}^{(n)}R^2 \\
\nonumber
&=& {}^{(n)}C_{ijkl}~{}^{(n)}C^{ijkl} - 4~{n-3 \over n-2}~\bigg(
{}^{(n)}R_{ab}~{}^{(n)}R^{ab} - {n \over 4(n-1)}~{}^{(n)}R^2\bigg).
\een
In equation (\ref{lgb}), by ${}^{(n)}C_{ijkl}$ one denotes the $n$-dimensional
Weyl tensor.\\
To proceed further, let us examine the limit of the above 
relation in asymptotically flat spacetime near the spatial infinity, i.e., when
$r \rightarrow \infty$. 
One commences with the limit behaviour of $n$-dimensional Ricci curvature tensor ${}^{(n)}R_{ij}$. Namely, taking 
into account
relation (\ref{eq1}) and the asymptotic limits given by (\ref{lim})-(\ref{lim1}), one arrives 
at the following expression:
\be
{}^{(n)}\na^{i}V~{}^{(n)}R_{ij} \simeq \cO \bigg({x_j \over r^{2(n-1)}}\bigg) + \cO(\alpha).
\ee
Next, the right-hand side of equation (\ref{sss}) will be provided by the limits of the forms
\ben
{}^{(n)}R_{ij}~{}^{(n)}R^{ij} &\simeq& \cO \bigg({1 \over r^{2(n-1)}}\bigg) + \cO(\alpha),\\
{}^{(n)}R^2 &\simeq& \cO \bigg({1 \over r^{2(n-2)}}\bigg).
\een
Restricting our considerations to the case when slices are conformally flat, i.e.,
${}^{(n)}C_{ijkl} = 0$, one reaches to the expressions
\ben
{}^{(n)}\cL_{GB} &\simeq& \cO \bigg({1 \over r^{2(n-1)}}\bigg),\\
{}^{(n)}\na_{a}\psi ~{}^{(n)}\na^{a}\psi &\simeq& \cO \bigg({1 \over r^{2(n-1)}}\bigg),
\een
we arrive at the conclusion that the volume integral can be rewritten in terms of the 
surface integral at the spatial
infinity. It can be readily verify that it implies
\be
\int_{vol} dV ~{}^{(n)}\na^{i}~\bigg( {}^{(n)}\na_{i} V + \alpha~\cO \bigg({1 \over r^{2n-1}}\bigg) \bigg)
= \int_{S_{\infty}}dS^i~\bigg( {}^{(n)}\na_{i} V + \alpha~\cO \bigg({1 \over r^{2n-1}}\bigg) \bigg).
\ee
One can easily see that the integrands do not influence on the value of the integral in question 
and ${}^{(n)}\na^{i}~{}^{(n)}\na_{i} V = 0$ which leads to the conclusion that 
$V$ is constant and in the spacetime under consideration the ADM mass is equal to zero.
\par
On this account it is customary to ask about the behaviour of the Ricci scalar curvature
on the considered conformally flat hypersurfaces. The sign of the Ricci scalar tensor plays the crucial
role in the rigid positive energy theorem \cite{posth}. Returning to the exact form of it which can be provided
by
\be
{}^{(n)}R = {2 \over V}~{}^{(n)}\na_{i}{}^{(n)}\na^{i} V - {2 \over n-1}T - \alpha~\bigg(
{n-3 \over n-1}~\bigg)~{}^{(n)}\cL_{GB},
\ee
one can reveal that
\be
{}^{(n)}R = 2~{{}^{(n)}\na_{i}\psi~{}^{(n)}\na^{i}\psi \over V^2}
~\bigg[
1 - \alpha~{4n~(n-3) \over (n-1)(n-2)}~{{}^{(n)}\na_{i}\psi~{}^{(n)}\na^{i}\psi \over V^2}
\bigg]
+ 4~\alpha~{n-3 \over n-2}~{}^{(n)}R_{ij}~{}^{(n)}R^{ij} + \cO(\alpha^2).
\ee
The second term is manifestly greater than zero, and because of the fact that we shall
restrict our consideration to $\cO(\alpha)$-order, one reveals that ${}^{(n)}R \geq 0$. Taking the above 
discussion into account, we can assert that the following theorem holds:\\
{\bf Theorem:}\\
Assume that one considers the asymptotically flat strictly static spacetime being
the solution of Einstein-Gauss-Bonnet equations of motion with $U(1)$-gauge matter sector.
Let us suppose further that the fall-off of the considered matter field in of the form
given by the relations (\ref{lim})-(\ref{lim1}). Restricting our considerations to the case of static conformally
flat slices of the spacetime in question, one states that up to the order of $\cO(\alpha)$ the spacetime is the
Minkowski one, where $\alpha$ is a coupling constant of the Gauss-Bonnet curvature correction.

\section{Black holes in Gauss-Bonnet gravity}
In this section we shall consider static black hole spacetime in the theory under consideration. 
The properties of black hole solutions of Einstein-Gauss-Bonnet gravity were 
intensively examined from various points of view. The thermodynamics of such black holes 
was investigated in Refs.\cite{neu04}, the spherical symmetric null dust collapse or scalar field
in Gauss-Bonnet gravity was elucidated in Refs.\cite{mae06}, while Vaidya type solutions 
were studied in \cite{vai}. On the other hand, black hole solutions for Einstein-Gauss-Bonnet
gravity were first achieved in \cite{bou85,mye88}, while for a charged case in Ref.\cite{wil88}.
The generalization of the above researches to the Gauss-Bonnet theory with a cosmological constant were presented in
\cite{gbc}, while the spacetime structure of $n$-dimensional static solution was elaborated in Refs.\cite{mae05}.
The above investigations were strengthen to the case of dilatonic Einstein-Gauss-Bonnet theory \cite{guo08},
where asymptotically flat and asymptotically AdS black holes in various dimensions were studied.
For a complete history of black holes in higher order gravity theories see, e.g., \cite{cha09} and references therein. 
\par
To proceed further, let us assume that in the  spacetime in question one can find appropriate coordinate system
in which the asymptotic limits determined by the relations (\ref{lim}-\ref{lim1}) will be
provided. Moreover, we suppose that the black hole event horizon is a Killing horizon located
at the level surface $V=0$. We shall consider the non-degenerate black hole Killing horizon case.
In the next step, one examines the conformal transformation which implies
\be
\tg^{\pm}_{ij} = \Omega^2_{\pm}~g_{ij}.
\ee
The conformal factor is provided by
\be
\Omega_{\pm} = \bigg[ \bigg( {1 \pm V \over 2} \bigg)^2 - {C^2 \over 4}~\psi^2 \bigg]^{1 \over n-2},
\ee
where $C = [2(n-2)/(n-1)]^{1/2}$. In the next step
the standard procedure used in the proofs of the black hole uniqueness theorem can be implemented.
Namely, we obtain two manifolds $(\Sigma_{-},~\tg^{-}_{ij})$ and $(\Sigma_{+},~\tg^{+}_{ij})$. Next,
we paste $(\Sigma_{\pm},~\tg^{\pm}_{ij})$ along the surface $V=0$. This in turn enables one
to construct a complete regular hypersurface $\Sigma = \Sigma_- \cup \Sigma_+ \cup \{ p\}$. On
the other hand, if $(\Sigma,~\tg_{ij})$ 
is an asymptotically flat solution of the underlying
equations of motion with non-degenerate black hole event horizon, then our next step will be
to study the total gravitational mass on hyperspace $\Sigma$. In order to do it we shall inspect
the Ricci scalar curvature given on the hypersurface in question.
It leads us t the two manifolds $(\Sigma^{\pm},~\tg^{\pm}_{ij})$.\\
The Ricci scalar curvature tensor ${}^{(n)}\tR$ on the adequate hypersurfaces $\Sigma^{\pm}$ is of the form
\be
{\Omega^4_{\pm}~{}^{(n)}\tR \over \bigg[ \bigg( {1 \pm V \over 2} \bigg)^2 - {C^2 \over 4}~\psi^2 \bigg]^{2\beta-2}} 
=
{2 \over V^4}~\bigg[ {1 \over 4}~\bigg( 1 - V^2 - C^2~\psi^2 \bigg)~{}^{(n)}\na^{i} 
\psi + {\psi V \over 2}~{}^{(n)}\na^{i} V \bigg]^2
+ \alpha~\bigg[ \bigg( {1 \pm V \over 2} \bigg)^2 - {C^2 \over 4}~\psi^2 \bigg]^2~\Theta,
\label{sca}
\label{ricci}
\ee
where we have denoted by $\Theta$ the following:
\be
\Theta = {n-3 \over (n-1)(n-2)}~\bigg( 4~(n-1)~{}^{(n)}R_{ij}^2 - n~{}^{(n)}R^2 \bigg) + {}^{(n)}C_{ijkl}^2.
\label{pos}
\ee
In what follows we shall discuss the two cases of the black hole solutions
in Einsein-Gauss-Bonnet gravity, namely uncharged and charged ones.

\subsection{Uncharged black hole solution}
In the case in question 
one has $\psi =0$ in the above relations. From equation (\ref{pos}) we get 
that the crucial point is the sign of it. Because of the fact that ${}^{(n)}R_{ij}^2$ and
${}^{(n)}C_{ijkl}^2$ are manifestly greater than zero, we focus our attention on
the term $- n~{}^{(n)}R^2$. From Ref.\cite{shi13} one gets that it is equal to
\be
{}^{(n)}R = 4~\alpha~{n-3 \over n-2}~{}^{(n)}R_{ij}^2 + \cO(\alpha^2).
\ee
Having in mind the exact form of relation (\ref{sca}), we conclude that $- n~{}^{(n)}R^2$
is of order $\cO(\alpha^2)$ and up to the $\alpha$-order can be neglected. Hence, it reveals
that ${}^{(n)}\tR$ is greater or equal to zero and the positive energy theorem \cite{posth} can be implemented in our
proof. Following the method presented in the next subsection one can prove the uniqueness of uncharged
asymptotically flat black holes in Gauss-Bonnet gravity up to $\cO(\alpha)$ order.
The above arguments enable one to formulate the conclusion:\\
{\bf Theorem}:\\
Let us consider a static solution to Gauss-Bonnet gravity equation of motion with an
asymptotically timelike Killing vector field $k_\mu$. Suppose that the manifold in question
consists of a connected and simply connected spacelike hypersurface $\Sigma$ to which $k_\mu$
is orthogonal. Then, up to $\cO(\alpha)$, where $\alpha$
is a Gauss-Bonnet coupling constant, there exist a neighborhood of $\Sigma$ which
is diffeomorphic to an open set of $n+1$-dimensional Schwarzschid-Tangherlini black hole
solution.

\subsection{Charged black hole solution}
In the charged black hole case the situation is far much more involved.
Having in mind the positive mass theorem, one has that the hypersurface $\Sigma$ must be flat and additionally the relation
provided by
\be
\bigg( V^2 + C^2~\psi^2 - 1 \bigg)~{}^{(n)}\na^{i} 
\psi  = 2~{\psi~ V}~{}^{(n)}\na^{i} V,
\ee
should be satisfied. The above equation yields that the level surface $V$ and the potential 
of $U(1)$-gauge field $\psi$ coincide, i.e., the physical Cauchy 
hypersurface $\Sigma$ is conformally flat.
\par
As far as the term bounded with $\Theta$ is concerned, after tedious computations one obtains 
\be
\Theta = {}^{(n)}C_{ijkl}^2 + 4~{n-1 \over V^2}~\bigg(
{}^{(n)}\na_{i}{}^{(n)}\na_{j} V - 2~{{}^{(n)}\na_{i}\psi {}^{(n)}\na^{i} \psi \over V} \bigg)^2
+ {n-3 \over (n-1)(n-2)}~W(n)~{({}^{(n)}\na_{i}\psi {}^{(n)}\na^{i} \psi)^2 \over V^4}
+ \cO(\alpha),
\ee
where we have written
\be
W(n) = {16(n-2)^2 + n~(n-1)^2 - 2~(n-1)~(n-2)~(n-3) - 3~n~(n-3)^2 \over (n-1)^2}.
\ee
A close inspection of the above polynomial reveals that for the spacetime dimensions equal to $5,~6,~7$
its value is greater than zero but when $n \geq 8$ is less than zero.
\par
As in the previous case we confine our attention to the $\cO(\alpha)$-order and use the positive energy theorem.
It is clear from equation (\ref{ricci}) that if one 
takes into account the small curvature corrections, the Ricci scalar curvature tensor ${}^{(n)}\tR(\tg)$ on the hypersurface
$\Sigma$ is equal to non-negative value. Moreover, using the asymptotic behaviour of $\tg_{ij +}$ on $\Sigma_{+}$
,i.e., $\tg_{ij +} = \delta_{ij} + {\cal O}\bigg( {1 \over  r^{n - 1}} \bigg) $, one concludes
that the total mass on the hypersurface $\Sigma$ vanishes. Consequently,
having in mind the positive mass theorem \cite{posth}, the manifold $\Sigma$ is isometric
to a flat manifold. This fact enables us to rewrite the metric tensor $g_{ij}$ in a conformally
flat form provided by the following expression \cite{gib02a}
\be
g_{ij} = {\cal U}^{1 \over n-2} \delta_{ij}.
\label{gg}
\ee
In the above relation we have defined a smooth function ${\cal U} = {2 \over 1 + V + C\phi}$
on the manifold $(\Sigma, \delta_{ij})$. 
It turns out that our equations of motion in the considered limit
reduce to the Laplace equation on the $n$-dimensional Euclidean manifold 
$
\na_{i}\na^{i}{\cal U} = 0,
$
where $\na$ is the connection on a flat manifold. 
Consequently one can endorse for the metric $\delta_{ij}$
in the flat base space the line element provided by
\be 
\delta_{ij} dx^{i}dx^{j} = \trho^{2} d{\cal U}^2 + {\tilde h}_{AB}dx^{a}dx^{B}.
\ee
To begin with we shall examine the case of the single black hole event horizon.
The event horizon is located at ${\cal U} = 2$ and
it can be shown that the embedding of $\cal H$ into the Euclidean
$n$-dimensional space is totally umbilical \cite{kob69}. Moreover, the embedding must be 
hyperspherical, which means that 
each of the connected components of the horizon $\cal H$
is a geometric sphere of a certain radius determined by the value of
$\rho \mid_{\cal H}$,
where $\rho$ is the coordinate introduced on the hypersurface $\Sigma$ by the line element
of the form
$$g_{ij}dx^{i}dx^{j} = \rho^2 dV^2 + h_{AB}dx^{A}dx^{B}.$$
Without loss of generality, one 
connected component of the horizon can be always located
at $r = r_{0}$ surface.
Thus, we have to do with a boundary value problem for the Laplace equation 
on the base space $\Omega = E^{n}/B^{n}$ with the
Dirichlet boundary condition ${\cal U} \mid_{\cal H} = 2$ and the asymptotic
decay condition ${\cal U} = 1 + {\cal O} \bigg( {1 \over r^{n-2}} \bigg)$.
Let us suppose further, that
${\cal U}_{1}$ and ${\cal U}_{2}$ be two solutions subject to the boundary value problem.
Using the Green identity and integration over the volume element
one arrives at the following expression:
\be
\bigg( \int_{r \rightarrow \infty} - \int_{\cal H} \bigg) 
\bigg( {\cal U}_{1} - {\cal U}_{2} \bigg) {\p \over \p r}
\bigg( {\cal U}_{1} - {\cal U}_{2} \bigg) dS = \int_{\Omega}
\mid \na \bigg( {\cal U}_{1} - {\cal U}_{2} \bigg) \mid^{2} d\Omega.
\ee
The boundary conditions make the left-hand side of the above relation vanish
and we conclude that two solutions must be identical.
\par
In order to get rid of the assumption about a single event horizon, one should elaborate
he evolution level surface in Euclidean space \cite{gib02a,gib02b}.
The Gauss equation in Euclidean space allows us to receive the evolution relations
concerning the behaviour of shear
$\sigma_{AB}$. Then, the harmonicity of $\cal U$ function leads us to the conclusion that
\be
\sigma_{AB} = 0, \qquad \hat {\cal D}_{A} \rho = 0, \qquad \hat {\cal D}_{A} k = 0,
\ee
where $\hat {\cal D}_{A}$ denotes the covariant derivative on each level set of $V$,
$k_{AB}$ is the second fundamental form of the level set. 
Consequently, this effects that each level surface of the function $\cal U$ is 
totally umbilic and therefore spherically symmetric.
\par
It worth mentioning that the above results are local \cite{gib02a,gib02b},
because of the fact that we have examined only the region without saddle points of the harmonic function
in question. As was claimed in the above references the global result 
can be achieved by taking into account the
assumption about analyticity.
Summing it all up, we arrive at the main statement of this subsection:\\
{\bf Theorem}:\\
Consider a static solution to $(n+)1$-dimensional Gauss-Bonnet gravity with an Abelian $U(1)$-gauge
field equations of motion. Suppose that there is defined an asymptotically
timelike Killing vector field $k_\mu$ orthogonal to the simply connected spacelike hypersurface
$\Sigma$. Then, up to the $\cO(\alpha)$-order, where $\alpha$
is a Gauss-Bonnet coupling constant,
it turns out that there is a neighborhood of the hypersurface in question
which is diffeomorphic to an open set of $(n+1)$-dimensional Reissner-Nordstr\"om non-extreme black hole
solution.

\section{Conclusions}
In our paper we have elaborated the strictly static asymptotically flat spacetimes
in Einstein-Gauss-Bonnet gravity theory with $U(1)$-gauge matter field. It was found that
up to the small curvature corrections and the adequate decay property
of the matter field, static 
conformally flat slices of the manifold in question
were of Minkowski origin. In the same limit we analyzed static black hole solutions, both
uncharged and charged ones. It was revealed that in the considered spacetime there were
neighborhoods which were diffeomorphic to $(n+1)$-dimensional Schwarzschild-Tangherlini and
non-extreme Reissner-Nordstr\"om black hole spacetimes. 
\par
Of course there are many remaining issues like, e.g, getting rid of the $\cO(\alpha)$-order limit or to consider
another type of black hole, stationary axisymmetric solution, in the theory in question.
We hope to consider these problems elsewhere.


\begin{acknowledgments}
MR was partially supported by the grant of the National Science Center
$2011/01/B/ST2/00488$.
\end{acknowledgments}




\begin{thebibliography}{99}
%
\def\cmp#1#2#3{{ Commun. Math. Phys.} {\bf #1}, #2 (#3)}
\def\lmp#1#2#3{{ Lett. Math. Phys.} {\bf #1}, #2 (#3)}
\def\cpam#1#2#3{{ Commun. Pure Appl. Math.} {\bf #1}, #2 (#3)}
\def\hpa#1#2#3{{ Hell. Phys. Acta} {\bf #1}, #2 (#3)}
\def\grg#1#2#3{{ Gen. Rel. Grav.} {\bf #1}, #2 (#3)}
\def\pr#1#2#3{{ Phys. Rev.} {\bf #1}, #2 (#3)}
\def\prl#1#2#3{{ Phys. Rev. Lett.} {\bf #1}, #2 (#3)}
\def\prd#1#2#3{{ Phys. Rev. D} {\bf #1}, #2 (#3)}
\def\pl#1#2#3{{ Phys. Lett} {\bf #1}, #2 (#3)}
\def\pla#1#2#3{{ Phys. Lett. A} {\bf #1}, #2 (#3)}
\def\plb#1#2#3{{ Phys. Lett. B} {\bf #1}, #2 (#3)}
\def\prep#1#2#3{{ Phys. Reports} {\bf #1}, #2 (#3)}
\def\phys#1#2#3{{ Physica} {\bf #1}, #2 (#3)}
\def\jcp#1#2#3{{ J. Comput. Phys.} {\bf #1}, #2 (#3)}
\def\jmp#1#2#3{{ J. Math. Phys.} {\bf #1}, #2 (#3)}
\def\jpm#1#2#3{{ J. Phys. A: Math. Gen.} {\bf #1}, #2 (#3)}
\def\jgp#1#2#3{{ J. Geom. Phys.} {\bf #1}, #2 (#3)}
\def\cpr#1#2#3{{ Computer Phys. Rept.} {\bf #1}, #2 (#3)}
\def\cqg#1#2#3{{ Class. Quantum Grav.} {\bf #1}, #2 (#3)}
\def\cma#1#2#3{{ Computers Math. Applic.} {\bf #1}, #2 (#3)}
\def\mc#1#2#3{{ Math. Compt.} {\bf #1}, #2 (#3)}
\def\apj#1#2#3{{ Astrophys. J.} {\bf #1}, #2 (#3)}
\def\apjs#1#2#3{{ Astrophys. J. Suppl.} {\bf #1}, #2 (#3)}
\def\acta#1#2#3{{ Acta Astronomica} {\bf #1}, #2 (#3)}
\def\apl#1#2#3{{Ann. Physik. (Leipzig)} {\bf #1}, #2 (#3)}
\def\anp#1#2#3{{Ann. Phys. } {\bf #1}, #2 (#3)}
\def\ahp#1#2#3{{Ann. Henri Poincare } {\bf #1}, #2 (#3)}
\def\ams#1#2#3{{Acta Math. Sin. } {\bf #1}, #2 (#3)}
\def\sa#1#2#3{{ Sov. Astro.} {\bf #1}, #2 (#3)}
\def\sia#1#2#3{{ SIAM J. Sci. Statist. Comput.} {\bf #1}, #2 (#3)}
\def\aa#1#2#3{{ Astron. Astrophys.} {\bf #1}, #2 (#3)}
\def\mnras#1#2#3{{ Mon. Not. R. astr. Soc.} {\bf #1}, #2 (#3)}
\def\npb#1#2#3{{ Nucl. Phys. B} {\bf #1}, #2 (#3)}
\def\prsla#1#2#3{{ Proc. R. Soc. London, Ser. A} {\bf #1}, #2 (#3)}
\def\jhep#1#2#3{{ JHEP} {\bf #1}, #2 (#3)}             
\def\nuc#1#2#3{{Nuovo Cimento B } {\bf #1}, #2 (#3)}
\def\ijmp#1#2#3{{Int. J. Mod. Phys. D} {\bf #1}, #2 (#3)}
\def\atmp#1#2#3{{Adv. Theor. Math. Phys.} {\bf #1}, #2 (#3)}
\def\ptps#1#2#3{{Prog. Theor. Phys. Suppl.} {\bf #1}, #2 (#3)}
\def\ptp#1#2#3{{Prog. Theor. Phys.} {\bf #1}, #2 (#3)}
\def\lmp#1#2#3{{Lett. Math. Phys. } {\bf #1}, #2 (#3)}
\def\mmj#1#2#3{{Mich. Math. j. } {\bf #1}, #2 (#3)}
\def\lnphys#1#2#3{{Lect. Notes Phys. } {\bf #1}, #2 (#3)}

\def\hepph#1#2{{ hep-ph }{\bf #1} (#2)}
\def\hepth#1#2{{ hep-th }{\bf #1} (#2)}
\def\grqc#1#2{{ gr-qc }{\bf #1} (#2)}
\def\ibid#1#2#3{{ {\it ibid.} }{\bf #1}, #2 (#3)}
%
\bibitem{isr}W.Israel, \pr{164}{1776}{1967}.
\bibitem{mil73}H.M\"uller zum Hagen, C.D.Robinson and H.J.Seifert,
\grg{4}{53}{1973},\\
H.M\"uller zum Hagen, C.D.Robinson and H.J.Seifert,
\grg{5}{61}{1974}.
\bibitem{rob77}C.D.Robinson, \grg{8}{695}{1977}. 
\bibitem{bun87}G.L.Bunting G.L and A.K.M.Masood-ul-Alam, 
\grg{19}{147}{1987}.
\bibitem{rub88}P.Ruback, \cqg{5}{L155}{1988}. 
\bibitem{ma1}A.K.M.Masood-ul-Alam, \cqg{9}{L53}{1992}.
\bibitem{heus}M.Heusler, \cqg{11}{L49}{1994},\\
M.Heusler, \cqg{10}{791}{1993}.
\bibitem{chr99}P.T.Chru\'sciel, \cqg{16}{661}{1999},\\
P.T.Chru\'sciel, \cqg{16}{689}{1999}.
\bibitem{chr06}
P.T.Chrusciel, H.S.Real, and P.Tod, \cqg{23}{549}{2006}.

\bibitem{stat}
B.Carter in {\it Black Holes}, edited
by C.DeWitt and B.S.DeWitt (Gordon and Breach, New York, 1973),\\
B.Carter in {\it Gravitation and Astrophysics}, edited
by B.Carter and J.B.Hartle (Plenum Press, New York, 1987),
C.D.Robinson, \prl{34}{905}{1975},
\bibitem{maz}P.O.Mazur, \jpm{15}{3173}{1982},\\
P.O.Mazur, \pla{100}{341}{1984}.
\bibitem{bun}G.L.Bunting, PHD thesis, Univ.of New 
England, Armidale N.S.W., 1983. 
\bibitem{book}P.O.Mazur,
{\it Black Hole Uniqueness Theorems}
\hepth{0101012}{2001},\\
M.Heusler, {\it Black Hole Uniqueness Theorems} 
(Cambridge: Cambridge University Press, 1997).

\bibitem{gib02a}
G.W.Gibbons, D.Ida, and T.Shiromizu, \prd{66}{044010}{2002}.
\bibitem{gib02b}
G.W.Gibbons, D.Ida, and T.Shiromizu, \prl{89}{041101}{2002}.

\bibitem{ndim}
S.Hollands, A.Ishibashi, and R.M.Wald, \cmp{271}{699}{2007},\\
M.Rogatko, \cqg{19}{L151}{2002},\\
M.Rogatko, \prd{67}{084025}{2003},\\
M.Rogatko, \ibid{70}{044023}{2004},\\
M.Rogatko, \ibid{71}{024031}{2005},\\
M.Rogatko, \ibid{73}{124027}{2006}.

\bibitem{nrot}
Y.Morisawa and D.Ida, \prd{69}{124005}{2004},\\
Y.Morisawa, S.Tomizawa, and Y.Yasui, \ibid{77}{064019}{2008},\\
M.Rogatko, \ibid{70}{084025}{2004},\\
M.Rogatko, \ibid{77}{124037}{2008},\\
S.Hollands and S.Yazadjiev, \cmp{283}{749}{2008},\\
S.Hollands and S.Yazadjiev, \cqg{25}{095010}{2008},\\
D.Ida, A.Ishibashi, and T.Shiromizu, \ptps{189}{52}{2011},\\
S.Hollands and A.Ishibashi, \cqg{29}{163001}{2012}.
\bibitem{rog12}
M.Rogatko, \prd{86}{064005}{2012}.

\bibitem{sugra}
A.K.M.Massod-ul-Alam, \cqg{14}{2649}{1993},\\
M.Mars and W.Simon, \atmp{6}{279}{2003},\\
M.Rogatko, \cqg{14}{2425}{1997},\\
M.Rogatko, \prd{58}{044011}{1998},\\
M.Rogatko, \ibid{59}{104010}{1999},\\
M.Rogatko, \ibid{82}{044017}{2010},\\
M.Rogatko, \cqg{19}{875}{2002},\\
S.Tomizawa, Y.Yasui, and A.Ishibashi, \prd{79}{124023}{2009},\\
S.Tomizawa, Y.Yasui, and A.Ishibashi, \prd{81}{084037}{2010},\\
J.B.Gutowski, \jhep{0408}{049}{2004},\\
J.P.Gauntlett, J.B.Gutowski, C.M.Hull, S.Pakis, and H.S.Real, \cqg{20}{4587}{2003},\\
M.Rogatko, \prd{89}{044020}{2014}.
\bibitem{shi13a}
T.Shiromizu and S.Ohashi, \prd{87}{087501}{2013}.

\bibitem{gre87}
M.B.Green, J.H.Schwarz, and E.Witten, {\it Superstring Theory}, vol.II (Cambridge University Press,
Cambridge, England, 1987).

\bibitem{shi13}
T.Shiromizu and K.Tanake, \prd{87}{081504}{2013}.

\bibitem{rog13}
M.Rogatko, \prd{88}{024051}{2013}.














\bibitem{adscft}
J.M.Maldacena, \atmp{2}{231}{1998},\\
S.S.Gubser, I.R.Klebanov, and A.M.Polyakov, \plb{428}{105}{1998}.

\bibitem{shi12}
T.Shiromizu, S.Ohashi, and R.Suzuki, \prd{86}{064041}{2012}.

\bibitem{bak13}
B.Bakon and M.Rogatko, \prd{87}{084065}{2013}.


\bibitem{br}
S.Nojiri and S.D.Odintsov, \jhep{07}{049}{2000},\\
S.D.Odintsov and S.Ogushi, \prd{65}{023521}{2002},\\
K.A.Meissner and M.Olechowski, \ibid{65}{064017}{2002}.

\bibitem{bou85}
D.G.Boulware and S.Deser, \prl{55}{2656}{1985}.

\bibitem{posth}
R.Schoen and S.T.Yau, \cmp{65}{45}{1979},\\
E.Witten, \ibid{80}{381}{1981}.


\bibitem{neu04}
Y.M.Cho and I.P.Neupane, \prd{66}{024044}{2002},\\
I.Neupane, \ibid{69}{084011}{2004}
\bibitem{mae06}
H.Maeda, \cqg{23}{2155}{2006},\\
T.Taves, C.D.Leonard, G.Kunstatter, and R.B.Mann, \cqg{29}{015012}{2012}.
\bibitem{vai}
T.Kobayashi, \grg{37}{1869}{2005},\\
A.E.Dominguez and E.Gallo, \prd{73}{064018}{2006},\\
S.G.Ghosh and D.W.Deshkar, \prd{77}{047504}{2008}.
\bibitem{mye88}
R.C.Myers and J.Z.Simon, \prd{38}{2434}{1988}.
\bibitem{wil88}
D.L.Wiltshire, \prd{38}{2445}{1988}.
\bibitem{gbc}
R.G.Cai, \prd{65}{084014}{2002},\\
R.G.Cai and Q.Guo, \ibid{69}{104025}{2004},\\
R.Aros, R.Troncoso, and J.Zanelli, \ibid{63}{084015}{2001},\\
M.Cvetic, S.Nojiri, and S.D.Odintsov, \npb{628}{295}{2002},\\
C.Charmousis, J.F.Dufaux, \cqg{19}{4671}{2002}.
\bibitem{mae05}
T.Tori and H.Maeda, \prd{71}{124002}{2005},\\
T.Tori and H.Maeda, \ibid{72}{064007}{2005}.
\bibitem{guo08}
Z.Guo, N.Ohta, and T.Tori, \ptp{120}{581}{2008},\\
Z.Guo, N.Ohta, and T.Tori, \ibid{121}{253}{2009},\\
K.Maeda, N.Ohta, and Y.Sasagawa, \prd{83}{044051}{2011},\\
N.Ohta and T.Torii, \ibid{86}{104016}{2012}.
\bibitem{cha09}
C.Charmousis, \lnphys{769}{299}{2009}.

\bibitem{kob69}
S.Kobayashi and K.Nomizu, {\it Foundation of Differential Geometry}, (Interscience Publishers, New York, 1969).






















\end{thebibliography}
\end{document}